\documentclass{optica-article}

\journal{opticajournal} 

\articletype{Research Article}

\usepackage{graphicx}
\usepackage{caption}
\usepackage{subcaption}
\usepackage{amsmath}
\usepackage{lineno}
\DeclareUnicodeCharacter{2009}{FIXME}

\begin{document}

\title{Chirped pulse waveguide amplifier}

\author{
Alexander Rudenkov,\authormark{1,*}
Vladimir L. Kalashnikov,\authormark{1}
Maxim Demesh,\authormark{1}
Nikolai Tolstik,\authormark{1}
Evgeni Sorokin,\authormark{2,3}
and Irina T. Sorokina\authormark{1,3}
}

\address{\authormark{1}Department of Physics, Norwegian University of Science and Technology, N-7491 Trondheim, Norway\\
\authormark{2}Photonics Institute, Vienna University of Technology, 1040 Vienna, Austria\\
\authormark{3}ATLA Lasers AS, Richard Birkelands vei 2B, 7034 Trondheim, Norway}

\email{\authormark{*}alexander.rudenkov@ntnu.no} 


\begin{abstract*} 
We introduce a single-mode Cr:ZnS crystalline waveguide ultrafast amplifier that provides a high gain of 5.5 dB/cm and 2.35 W of average output power. The depressed-cladding buried waveguide is produced by an ultrafast laser writing procedure, which allows a high degree of flexibility in fabrication when the geometry, size, and even effective index can be modified along the waveguide. An analytical model that includes both, pump and pulse propagation, allows calculation and optimization of the waveguide design. In a CPA arrangement with a volume Bragg grating-based stretcher/compressor, we demonstrate a broadband 34-mm long amplifier in a polycrystalline Cr:ZnS with a single-pass gain factor of 75 (5.5 dB/cm) oand a high average output power up to 2.35 W.  

\end{abstract*}

\section{Introduction}

The mid-infrared (mid-IR) wavelength range, which is often called a “molecular fingerprint” region, and in particular, the range between 2 and 5~$\mu$m is characterized by the presence of strong fundamental and overtone rovibrational absorption bands of atmospheric constituents, vapors and gases, such as carbon monoxide and dioxide, methane, ammonia, NOx, etc. The ability of ultrashort-puled solid-state lasers to spectrally cover the wide wavelength range, containing all the above bands simultaneously by an ultrashort pulse source or by rapid laser tuning is the main advantage of the ultrabroadband Cr$^{2+}$- II-VI compound based lasers \cite{DeLoach_Transition_metal_doped_zinc_chalcogenides, Page_Cr_doped_zinc_chalcogenides, SOROKINA_Cr_doped_II_VI_materials, Sorokina_Femtosecond_Cr_Based_Lasers, Mirov_Progress_2015}, making them particularly attractive for various environmental and climate change related sensing tasks \cite{Girard_Acetylene_2006, Sorokin_Sensitive_multiplex_2007, Bernhardt_dual_comb_2010}. 

Interestingly, Cr$^{2+}$-doped ZnS laser \cite{Sorokina_Broadly_tunable_2002, Sorokina_Continuous_wave_2002} may provide a path for direct integration with Si photonics, because ZnS is one of the few semiconductors lattice-matched to Si and because the Cr$^{2+}$:ZnS operation wavelength range of 2-3.1~$\mu$m \cite{SOROKINA_Cr_doped_II_VI_materials, sorokina2003solid} lies well in the transparency window of Si and allows for high-power broadest tunable and few-cycle ultrafast laser operation, including frequency combs \cite{Sorokina_Femtosecond_Cr_Based_Lasers, Sorokin_SSL_2013, Tolstik_SSL_2013, Mirov_Frontiers_2018} even in the directly diode-pumped setups \cite{Sorokina_Continuous_wave_2002, Nagl_PhD_Ths_2022}. The works on power scaling have enabled generation, via nonlinear frequency conversion, of mid-IR light that spans the entire “molecular fingerprint” region between 3 and 14~$\mu$m \cite{Mirov_Frontiers_2018, Nagl_PhD_Ths_2022}. For ultrashort-pulse Si photonics, it is also advantageous that Cr:ZnS wavelength range around 2-3~$\mu$m lies beyond the two-photon absorption range of Si below 2.1~$\mu$m.

The first step on this challenging but highly rewarding path of photonic integration was made twenty years ago in 2004, when two groups reported inscription of depressed cladding waveguides for future laser application in lithium fluoride with coilor-centers \cite{Kawamura_LiF_DFB_2004} and in Ti:sapphire crystal \cite{Apostolopoulos_First_TiS_WG_2004}. Immediately following these conceptual works A. Okhrimchuk et al. reported the first depressed cladding waveguide laser in a neodymium-doped YAG crystal \cite{Okhrimchuk:05}. Ten years later J.R. Macdonald et al. demonstrated waveguide structures based on polycrystalline Cr:ZnS \cite{Macdonald:14}. The tendency for the scattering losses to decrease with wavelength as 1/~$\lambda$$^{4}$ makes the waveguide laser writing even more attractiveg when applied to active media in the mid-IR range such as Cr:ZnS \cite{Macdonald:14, Tolstik:19, Sorokin:22}, Cr:ZnSe \cite{Macdonald:13, 10.1117/12.2079396, McDaniel:16, Berry:13}, Fe:ZnSe \cite{McDaniel:16, 10.1063/1.4927384} as well as Tm:ZBLAN \cite{Lancaster:11} and Ho:YAG \cite{Thorburn:17, McDaniel:17}. Further work on optimization of the waveguide parameters inevitably led to power-scaling in Yb:YAG up to 5.6 W output power \cite{Hakobyan:16} and in Cr:ZnSe laser it was scaled up to an impressive value of 5.1 W with a slope efficiency of 41\% \cite{McDaniel:16}, however, yet in a multimode regime.

The next important step has been made when N. Tolstik et al. reported the first single-mode single-crystal Cr:ZnS depressed cladding buried waveguide manufactured by femtosecond laser writing. The laser yielded 150 mW average power at 2272 nm wavelength with an 11\% slope efficiency \cite{Tolstik:19}. The most recent step towards photonics integration was made with demonstration of the first all-laser-microprocessed waveguide Cr:ZnS laser \cite{Sorokin:22}.
However, all these achievements have been made in lasers that are based on relatively short (within 1 cm length) waveguides, and no amplifier-waveguide setups have been reported so far. 
In the present work, we present the results of what is to our knowledge the first chirped-pulse amplification (CPA) \cite{STRICKLAND1985447} with depressed cladding waveguides that provide relatively high gain of up to 75 at average power levels of up to 2.35 W.

\section{Experimental setup}
\begin{figure}[ht!]
\centering\includegraphics[width=12.7cm]{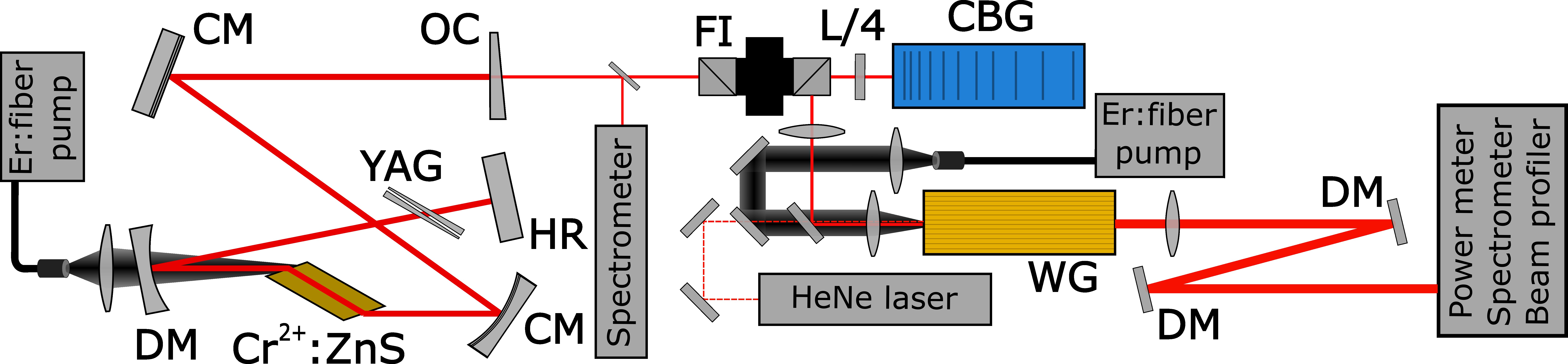}
\caption{Experimental setup: CM-chirped mirror, OC-output coupler, FI-Faraday isolator, L/4-quarter wave plate, CBG-chirped Bragg grating, YAG-wedges for dispersion adjustment, HR-highly reflective mirror, DM-dichroic mirror, WG- Cr$^{2+}$:ZnS polycrystalline active element with inscribed waveguides.}
\label{fig:fig_setup}
\end{figure}

To carry out the experiments we created an experimental setup, and its schematic is shown in Fig.~\ref{fig:fig_setup}.

\begin{figure}[ht!]
\centering\includegraphics[width=12.7cm]{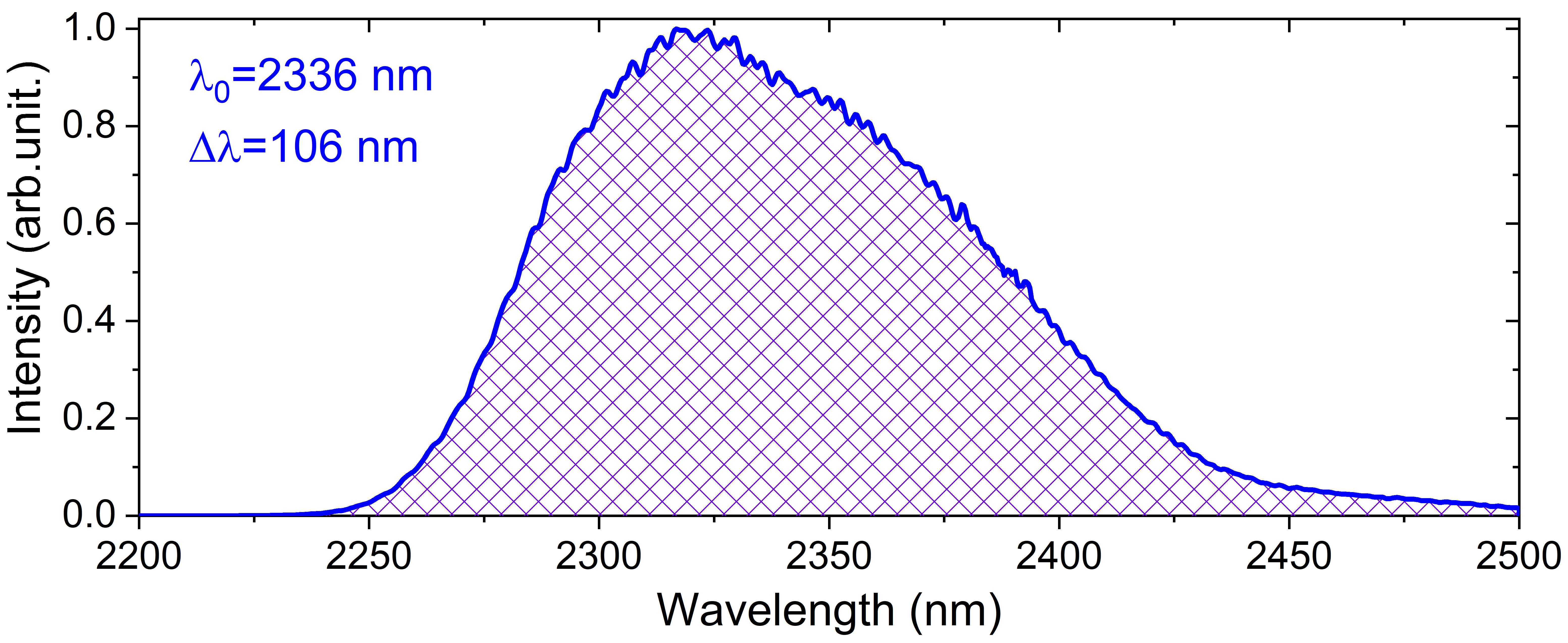}
\caption{Seed laser radiation spectrum.}
\label{fig:seed_laser_spectrum}
\end{figure}

As a seed pulse source we used the Kerr-lens mode-locked Cr$^{2+}$:ZnS laser (similar to initial 70 MHz laser described in \cite{Rudenkov:23}) pumped through dichroic mirrors (DM) by a low-noise 1610 nm Er:fiber laser. Intracavity dispersion compensation was achieved by broad-band chirped mirrors (CM).  Measured seed laser output radiation spectrum is illustrated in Fig.~\ref{fig:seed_laser_spectrum}. The FWHM spectral width of 106 nm at a central wavelength of 2336 nm corresponds to a sub-60 fs pulse duration. Such spectral parameters of seed laser radiation allowed a good overlap with the chirped volume Bragg grating (CBG) reflection band that will be illustrated in the experimental results block of the paper. A Faraday isolator was placed after the seed to eliminate any possible influence on the seed laser from the amplifier section.

In this experiment, we mainly focused on the study of the power/energy gain performance of the waveguides and used as seed pulses chirped ones provided by the specially designed CBG provided by OptiGrate. Typically, the performance of CBGs is limited by the trade-off between chirped pulse duration (dispersion), spectral bandwidth, and efficiency. In our case, about 500~ps pulse duration was provided by CBG with 100~nm spectral bandwidth centered at 2.35~$\mu$m. The reflection coefficient was about 50\%. Such a CBG can be successfully utilized as a stretcher unit, and enough long pulse duration eliminates most nonlinear effects that could occur during amplification. Next, laser beams from the seed, pump, and red guide laser were focused onto the input face of the active element. Seed laser pulses average power incident to the waveguide was about 100~mW.

For waveguide pumping, we used an unpolarized Er:fiber laser (IPG Photonics). A waveplate-polarizer beam attenuator provided 25 W maximum pump power at 1550~nm central wavelength, focused to about 18 $\mu$m at the waveguide entrance. For visual orientation, we used a red He-Ne laser beam to tune to a specific waveguide. The laser beams of the seed, pump, and red guide lasers have been adjusted coaxially over a large distance and ensured a fairly good visual preliminary coupling into the waveguide. Final optimization was then carried out using the amplified pulses average power at a reduced pump power. After the amplifier The collimating lens and dichroic pump/seed separator mirrors were placed at the active element output. An Ophir 30(150)A-BB-18 power meter, Yokogawa AQ6375 (1200-2600~nm) spectrometer, and Thorlabs BP109-IR2 slit scanning beam profiler (1000-2700~nm) were used to diagnose the amplified radiation parameters.

For the direct laser writing of the waveguides, we used an integrated picosecond Ho:YAG MOPA laser system (ATLA Lasers AS) operating at 2090 nm central wavelength.   
We fabricated 31 circular waveguides with different diameters of 20–50~$\mu$m that were inscribed in a 34-mm long polycrystalline Cr$^{2+}$:ZnS (Cr$^{2+}$ concentration of $2.9 \cdot 10$$^{18}$ cm$^{-3}$) sample using a NA=0.85 objective. Recording was carried out in the single-pulse modification mode (i.e., each defect is formed by one laser pulse without overlapping) with a step between defects (in the direction of the optical axis of the waveguide) of 3 and 4~$\mu$m. Such type of modification is equivalent to a free-form 3D printing and makes possible a great degree of flexibility in the waveguide design.
More detailed information on waveguides writing conditions can be found in \cite{demesh:000}.

\section{Experimental results}
During the experiment, the dependence of the amplified pulses' average power on the incident pump power for each waveguide was measured. To eliminate the influence of the uncoated faces of the active element and the coupling conditions, the gain factor was estimated by comparing the seed pulses' average powers measured at the same place at the waveguide output (after the separating mirrors) with and without applied pump power ($\Gamma$-coefficient in \ref{sec:model}). 

\begin{figure}[htbp]
    \centering
    \begin{minipage}[b]{0.7\textwidth}
        \centering
        \includegraphics[width=\linewidth]{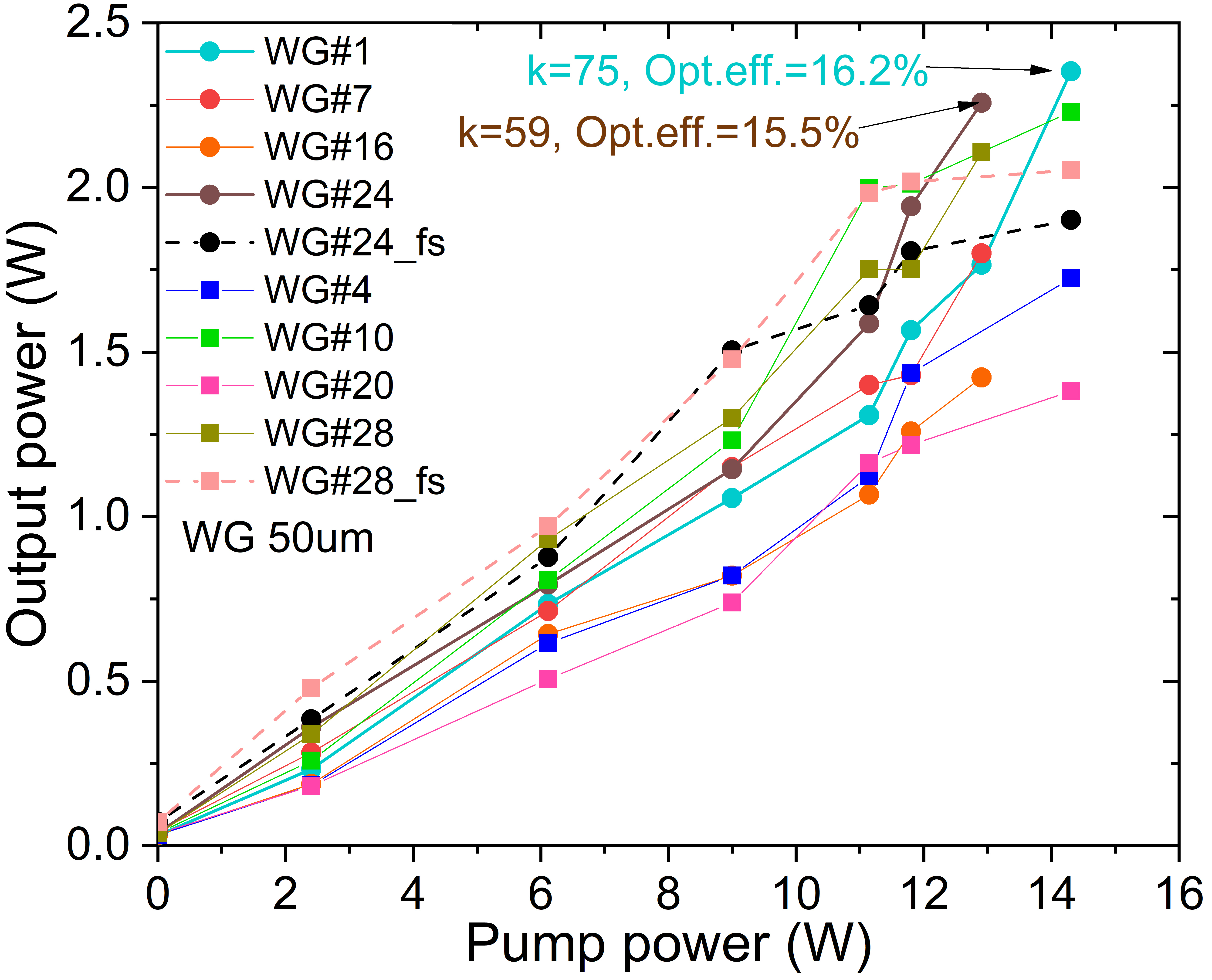} %
        \subcaption{Output power characteristics}
        \label{fig:50um_big_image}
    \end{minipage}
    \hfill
    \begin{minipage}[b]{0.27\textwidth}
        \centering
        \begin{subfigure}[b]{\textwidth}
            \centering
            \includegraphics[width=\linewidth]{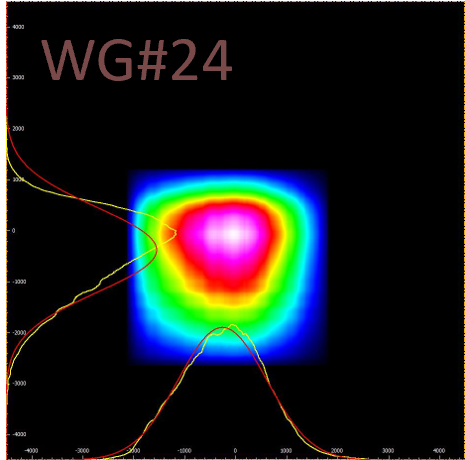} %
            \caption{WG\#24}
            \label{fig:50um_small_image1}
        \end{subfigure}
        \vfill
        \begin{subfigure}[b]{\textwidth}
            \centering
            \includegraphics[width=\linewidth]{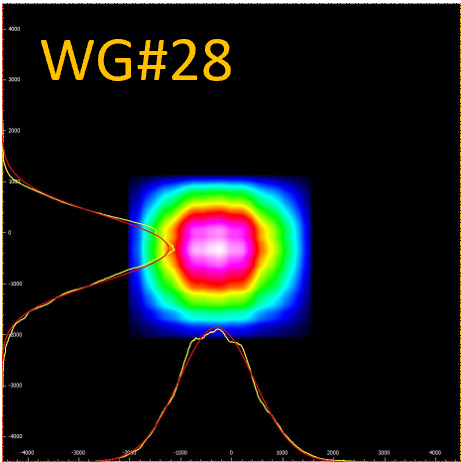} %
            \caption{WG\#28}
            \label{fig:50um_small_image2}
        \end{subfigure}
    \end{minipage}
    \caption{Output power characteristics (a) and beam profiles (b and c) of the 50~$\mu$m waveguides. In subfigure (a): the circles correspond to a defect recording step of 3~$\mu$m, and the squares --4~$\mu$m.}
    \label{fig:Output_power_50um}
\end{figure}

\begin{figure}[htbp]
    \centering
    \begin{minipage}[b]{0.46\textwidth}
        \centering
        \includegraphics[width=\linewidth]{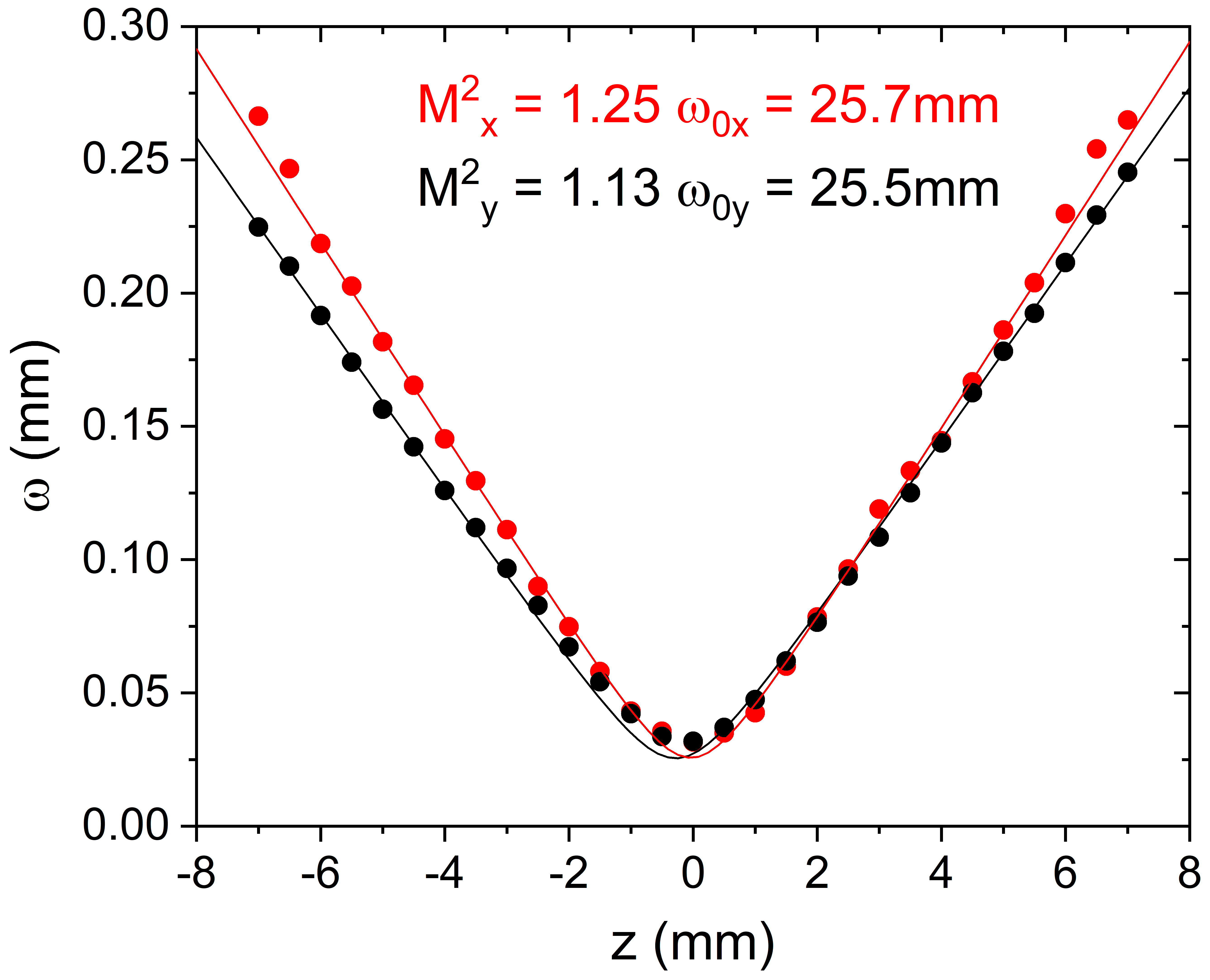} %
        \subcaption{M$^{2}$ factor measurement results.}
        \label{fig:M2_factor}
    \end{minipage}
    \hfill
    \centering
    \begin{minipage}[b]{0.45\textwidth}
        \centering
        \includegraphics[width=\linewidth]{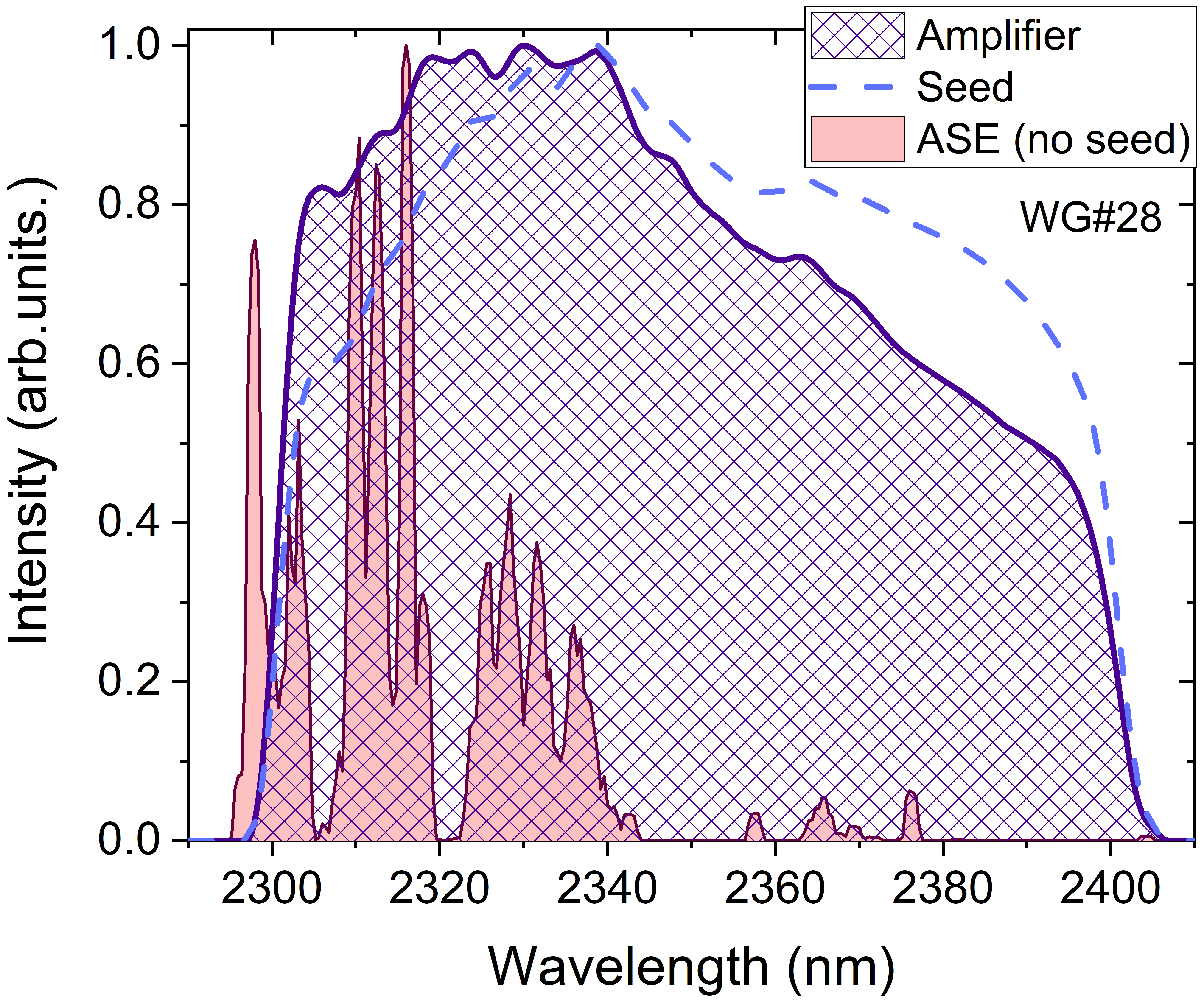} %
        \subcaption{Seed, amplifier, and ASE spectra.}
        \label{fig:wg_amp_spectra}
    \end{minipage}
\caption{WG\#28 M$^{2}$ factor (a) and seed, amplifier, and ASE spectra (b).}
\label{fig:M2_factor_wg_amp_spectra}
\end{figure}

The obtained experimental results are grouped into four groups according to the corresponding waveguide diameter. The best average output power, gain factor, and beam quality results were obtained with 50~$\mu$m waveguides and shown in Fig.~\ref{fig:Output_power_50um}. The maximum output power of 2.35~W and gain factor of 75 was obtained for waveguide WG\#1 with a power-added optical efficiency of 16.2\% under 14.3~W incident pump power. Waveguide WG\#28 demonstrated similar results (2.11/12.9~W output/incident power, gain factor of 57 and 16.1\% opt. eff.), and for this waveguide we observed M$^{2}$ factor of 1.13x1.25 (in vertical and horizontal planes, respectively) of the output beam (see Fig.~\ref{fig:M2_factor}). The seed amplified pulse and amplified spontaneous emotion (ASE) spectra of the WG\#28 waveguide are shown in Fig.~\ref{fig:wg_amp_spectra}.

The seed pulse spectrum (see Fig.~\ref{fig:wg_amp_spectra}) has a shape close to rectangular, which is a consequence of the spectrum edges cutting off upon reflection from the CBG. The effect of spectrum gain narrowing on the amplified pulses spectrum shape is also clearly visible: there is a predominant amplification of the short-wavelength region while the long-wavelength region has moderate amplification, which is also observed in the ASE spectrum, where the radiation is concentrated mainly in the short-wavelength region of the spectrum near the gain peak of the material.

\begin{figure}[htbp]
    \centering
    \begin{minipage}{0.49\textwidth}
        \centering
        \includegraphics[width=\linewidth]{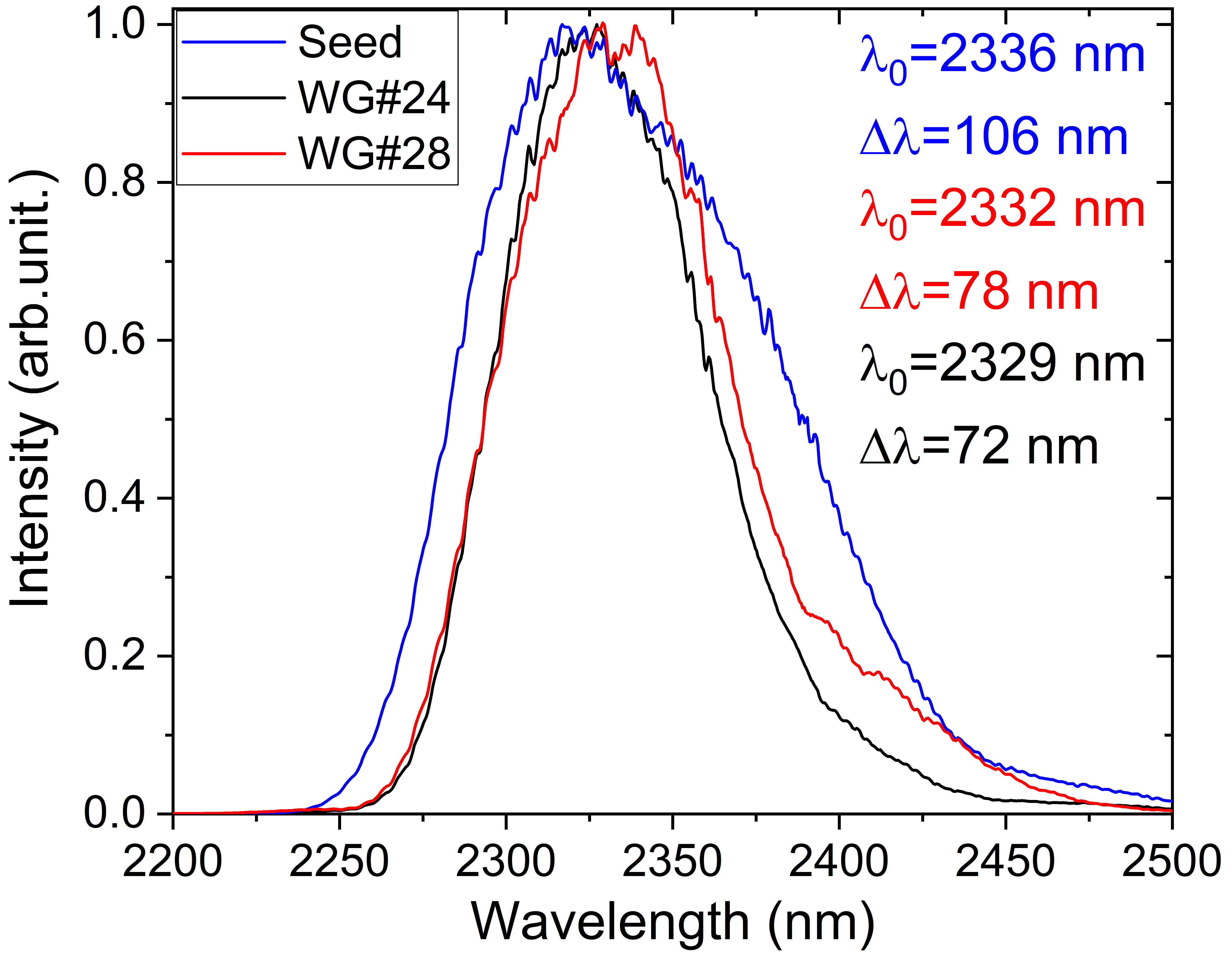}
        \subcaption{Seed and amplified pulse spectra.}
        \label{fig:fs_wg_amp_spectra}
    \end{minipage}
    \hfill
    \begin{minipage}{0.49\textwidth}
        \centering
        \includegraphics[width=\linewidth]{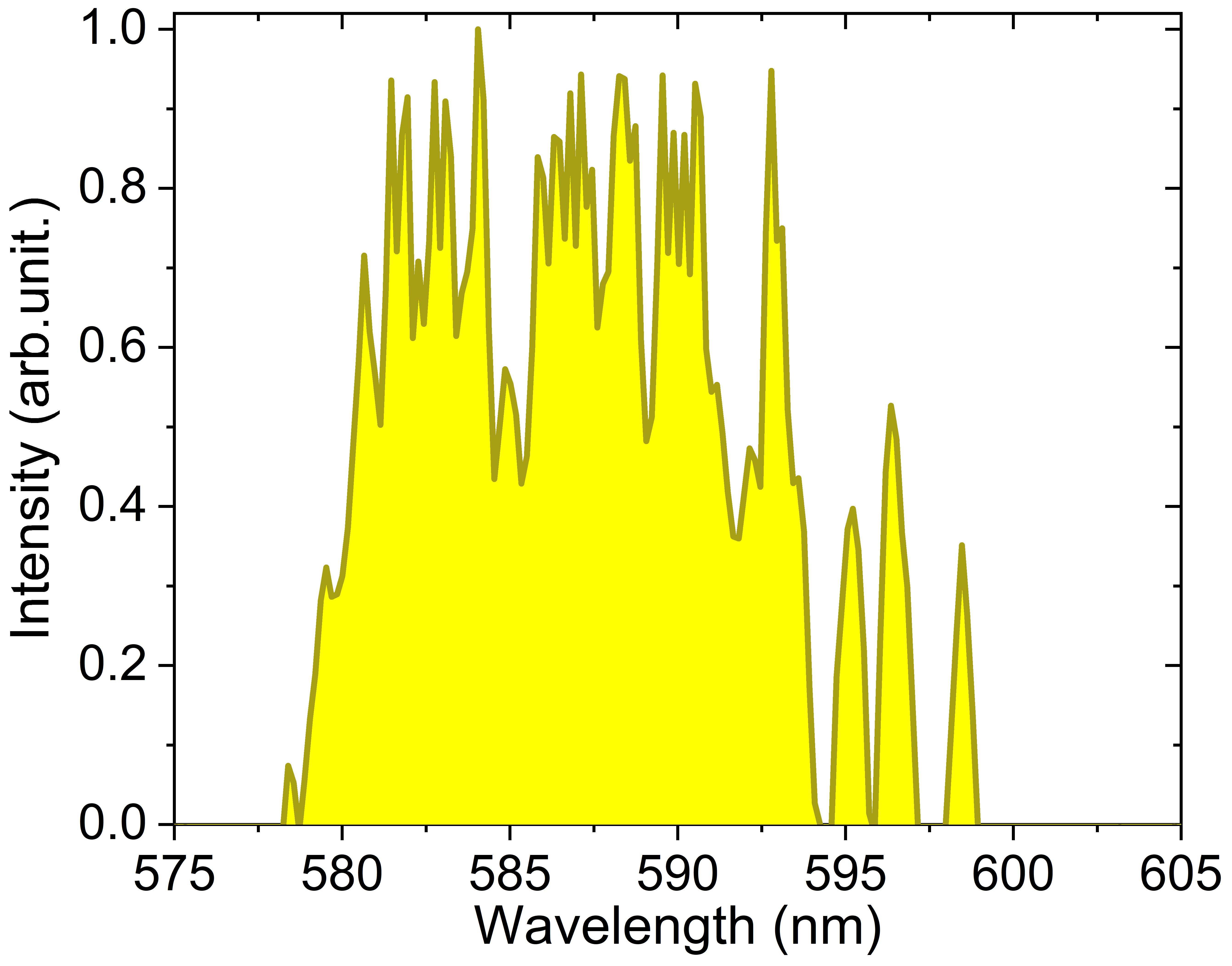}
        \subcaption{4th harmonic spectrum.}
        \label{fig:fs_wg_amp_broad_spectrum}
    \end{minipage}

    \vfill
 \vspace{0.5cm} %
    
    \begin{minipage}{0.49\textwidth}
        \centering
        \includegraphics[width=0.8\linewidth]{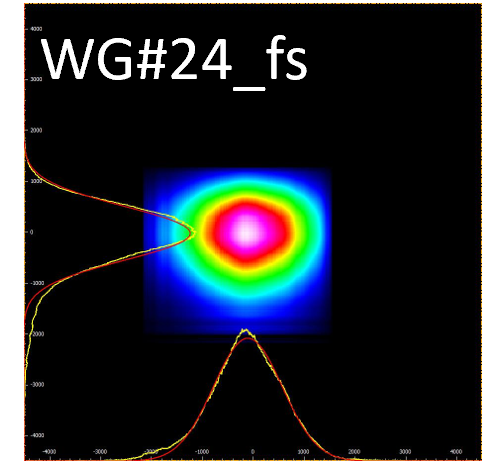}
        \subcaption{Output beam profile.}
        \label{fig:WG24_fs}
    \end{minipage}
    \hfill
    \begin{minipage}{0.49\textwidth}
        \centering
        \includegraphics[width=0.74\linewidth]{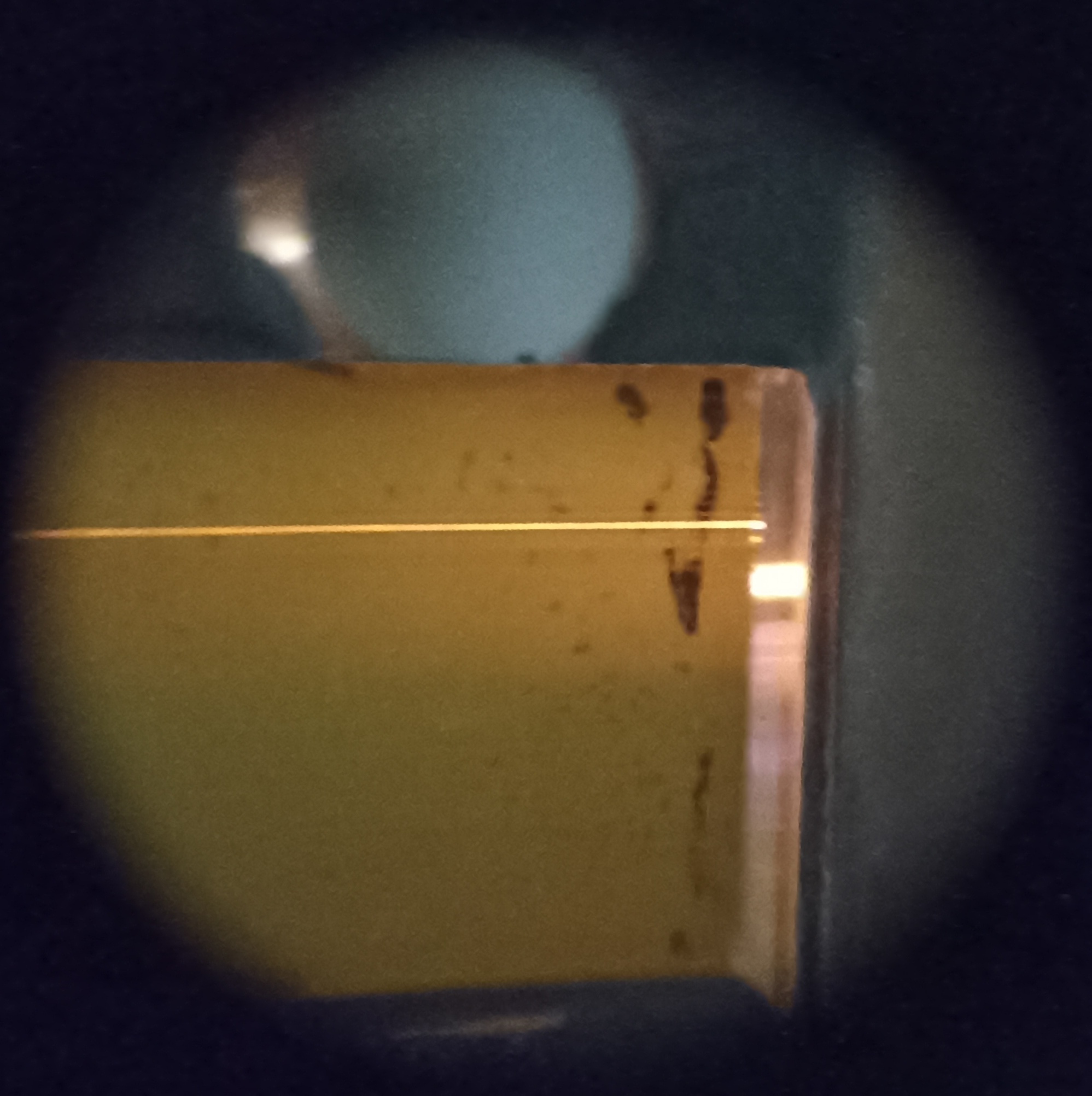}
        \subcaption{4th harmonic generation in waveguide.}
        \label{fig:visible_waveguide}
    \end{minipage}
    
    \caption{Waveguide amplifier in sub-ps regime: Seed, amplified pulse spectra (a), 4th harmonic spectrum (b), WG\#24 output beam profile (c), and photo of the 4th harmonic generation in waveguide (d). 
    }
    \label{fig:fs_spectra_beam_photo}
\end{figure}

Fig.~\ref{fig:50um_big_image} also shows WG\#24\_fs and WG\#28\_fs curves, corresponding to an experiment with amplification of pulses without its temporal stretching. We installed a highly reflective mirror in front of the CBG for this experiment. Thus, the pulse duration was in the sub-picosecond range. The main goal of this experiment was to test the hypothesis about the influence of nonlinearity on the amplified pulse beam forming, as well as preliminary experiments on nonlinear spectral broadening of amplified pulses. Fig.~\ref{fig:fs_spectra_beam_photo} illustrates waveguide amplifier spectra, 4th harmonic, and output beam profile in sub-ps mode. Typical gain-narrowing behavior is observed at the fundamental harmonic spectral range (see Fig.~\ref{fig:fs_wg_amp_spectra}) for WG\#24 and WG\#28 waveguides. The nonlinear conversion products (up to 4th harmonic generation) are noticeable over a wide spectrum in the case of sub-ps amplified pulse durations (see Fig.~\ref{fig:fs_spectra_beam_photo} (a) and (d)). The high frequency harmonics provide ready-to-use tools for CEP-stabilization of the whole setup \cite{MirovCEO}.

It should also be noted that the output beam profile remains almost unchanged within the entire pump power range, indicating the absence of the nonlinear beam cleaning effect at our experimental conditions.    

\begin{figure}[ht!]
\centering\includegraphics[width=7cm]{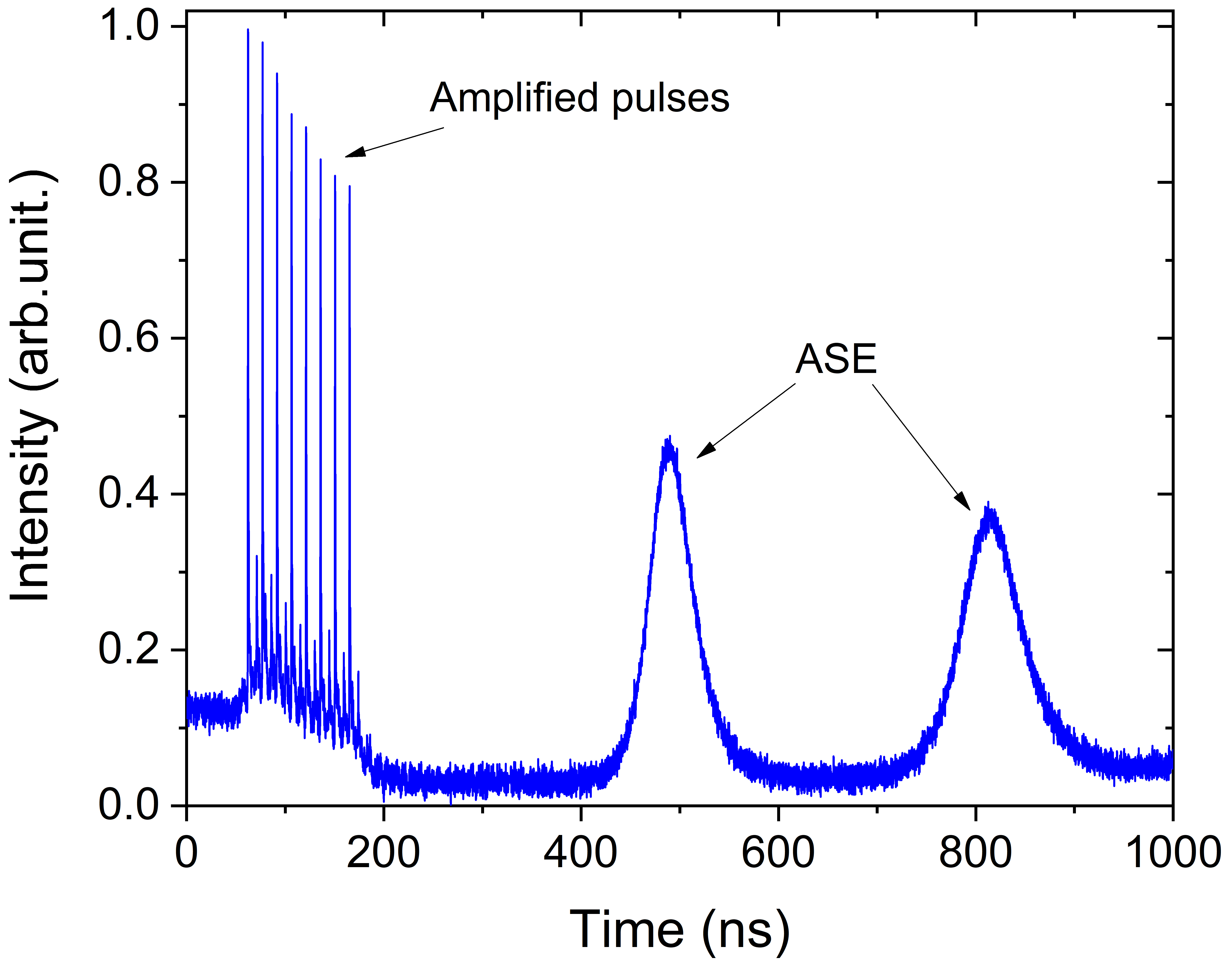}
\caption{Temporal radiation profile at the waveguide output in burst mode.}
\label{fig:burst_mode}
\end{figure}

Due to the demonstrated high gain of the 50~$\mu$m waveguides, we tested it in burst mode of operation. For this purpose, a Pockels cell (PC) and a polarizer were introduced into the experimental setup in front of the waveguides. The KTP-based PC operated at a pulse repetition frequency of 100~kHz and produced bursts of about 150~ns duration, including 8 seed pulses. Fig.~\ref{fig:burst_mode} shows the radiation's time profile at the waveguide amplifier's output. The presence of an amplified pulse burst is clearly noticeable, as are the initial  pulses of the of amplified spontaneous emission recovery.

\begin{figure}[htbp]
    \centering
    \begin{minipage}[b]{0.7\textwidth}
        \centering
        \includegraphics[width=\linewidth]{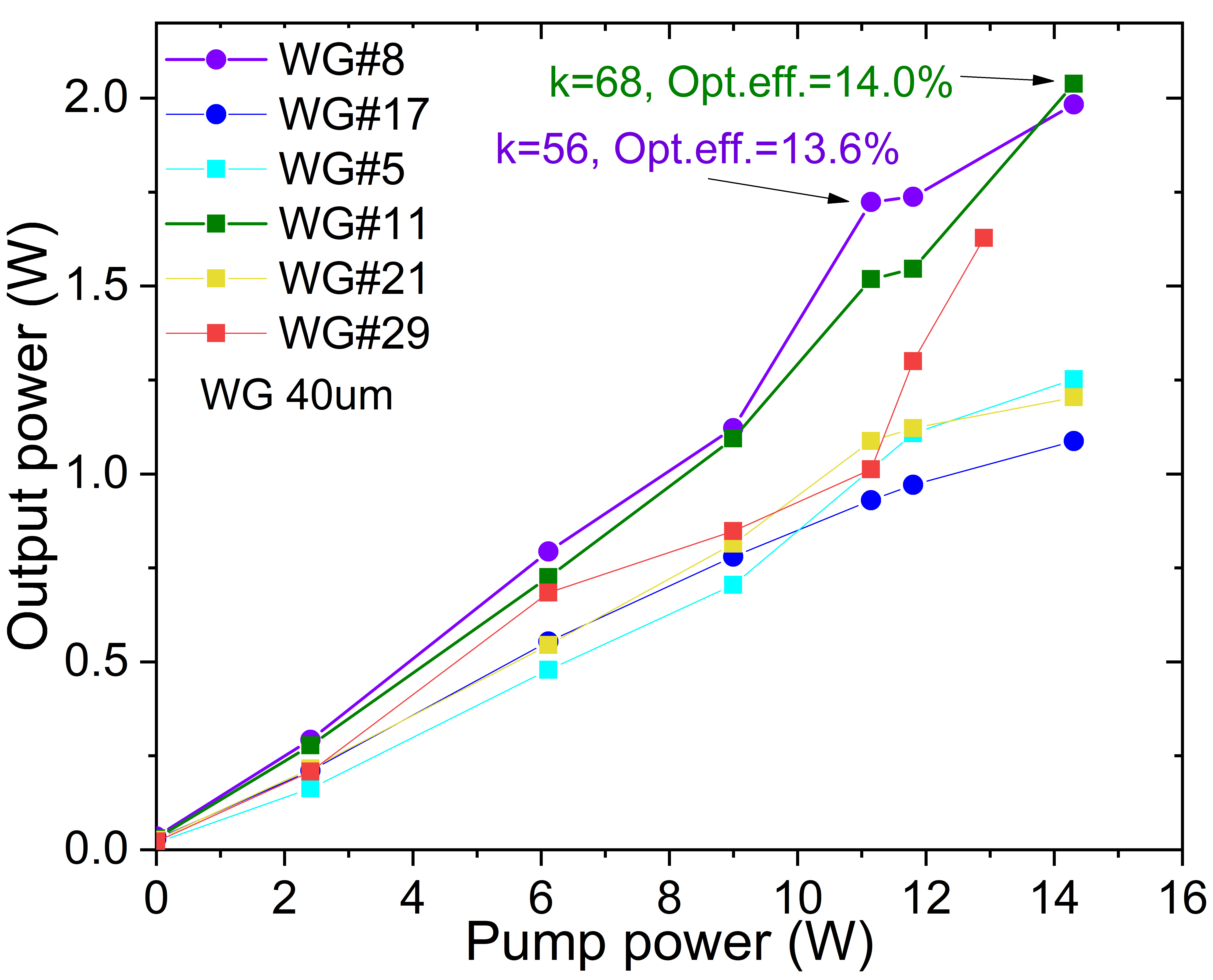} %
        \subcaption{Output power characteristics}
        \label{fig:40um_big_image}
    \end{minipage}
    \hfill
    \begin{minipage}[b]{0.27\textwidth}
        \centering
        \begin{subfigure}[b]{\textwidth}
            \centering
            \includegraphics[width=\linewidth]{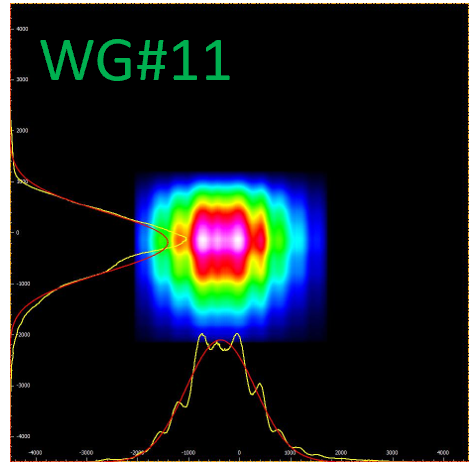} %
            \caption{WG\#11}
            \label{fig:40um_small_image1}
        \end{subfigure}
        \vfill
        \begin{subfigure}[b]{\textwidth}
            \centering
            \includegraphics[width=\linewidth]{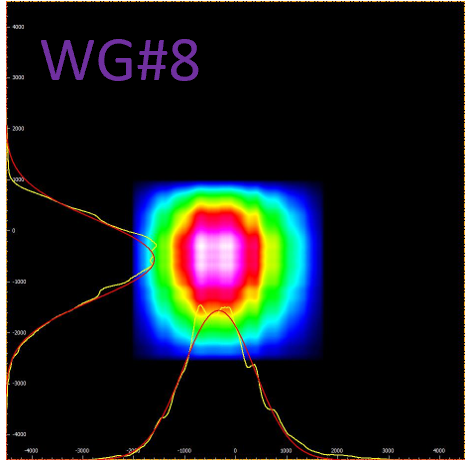} %
            \caption{WG\#8}
            \label{fig:40um_small_image2}
        \end{subfigure}
    \end{minipage}
    \caption{Output power characteristics (a) and beam profiles (b and c) of the 40~$\mu$m waveguides. In subfigure (a): the circles correspond to a defect recording step of 3~$\mu$m, and the squares-4~$\mu$m.}
    \label{fig:Output_power_40um}
\end{figure}

Waveguides with a diameter of 40~$\mu$m showed slightly lower, but overall quite similar to 50~$\mu$m ones results (see Fig.~\ref{fig:Output_power_40um}). The maximum output power of 2.04~W and gain factor of 68 was obtained for waveguide WG\#11 with a power-added optical efficiency of 14.0\% under 14.3~W incident pump power. 

\begin{figure}[htbp]
    \centering
    \begin{minipage}[b]{0.7\textwidth}
        \centering
        \includegraphics[width=\linewidth]{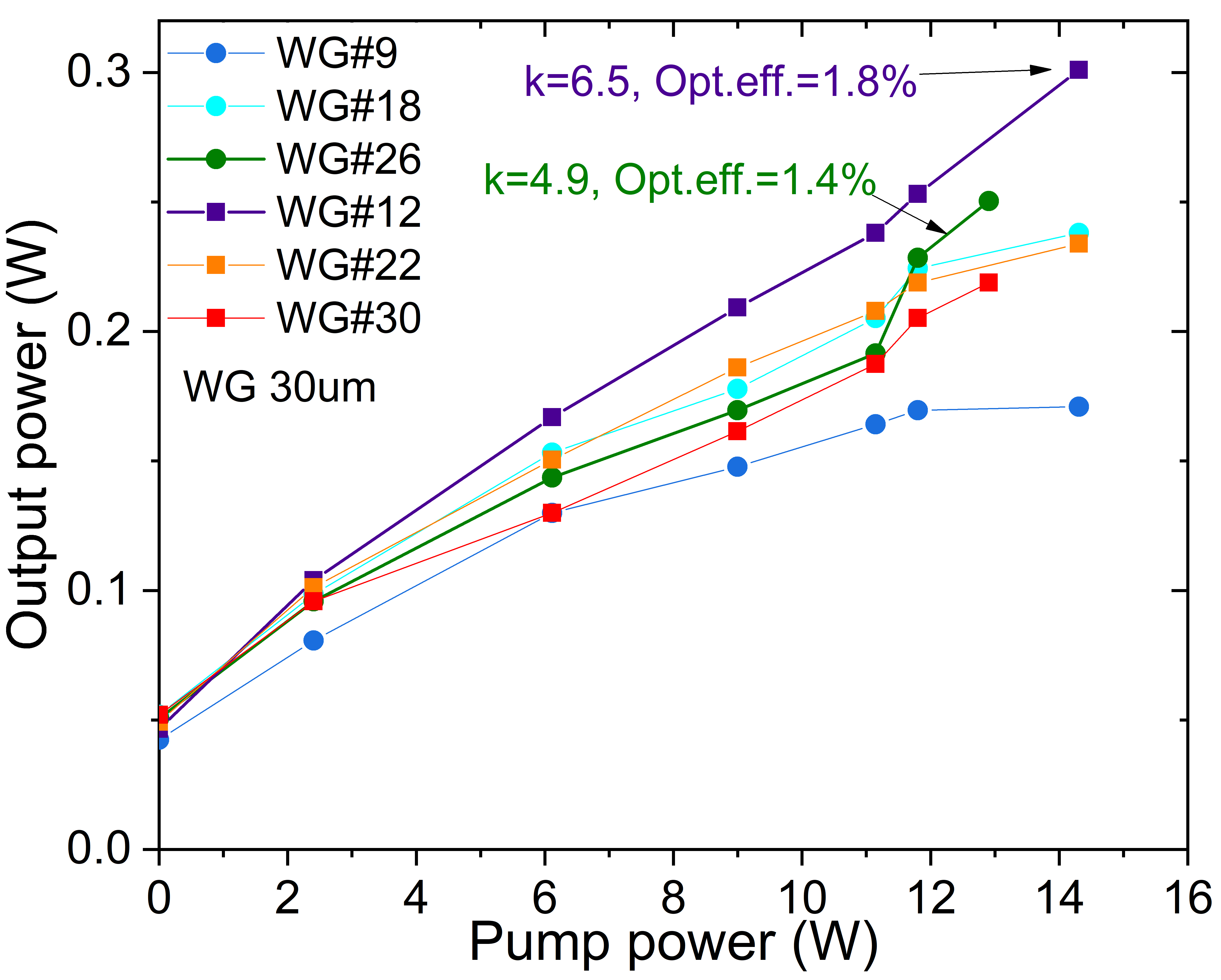} %
        \subcaption{Output power characteristics}
        \label{fig:30um_big_image}
    \end{minipage}
    \hfill
    \begin{minipage}[b]{0.27\textwidth}
        \centering
        \begin{subfigure}[b]{\textwidth}
            \centering
            \includegraphics[width=\linewidth]{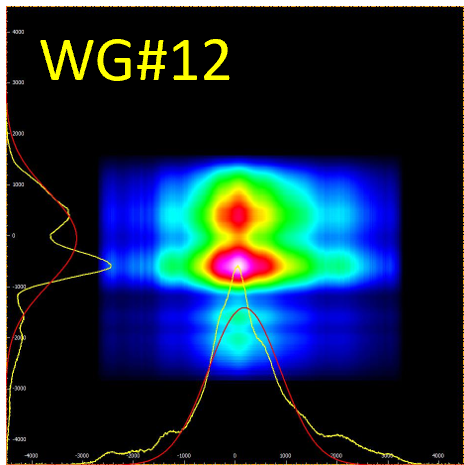} %
            \caption{WG\#12}
            \label{fig:30um_small_image1}
        \end{subfigure}
        \vfill
        \begin{subfigure}[b]{\textwidth}
            \centering
            \includegraphics[width=\linewidth]{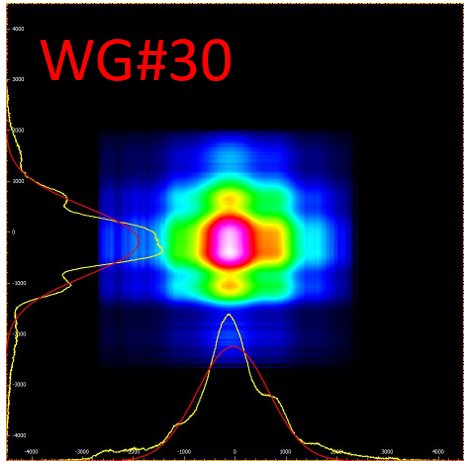} %
            \caption{WG\#30}
            \label{fig:30um_small_image2}
        \end{subfigure}
    \end{minipage}
    \caption{Output power characteristics (a) and beam profiles (b and c) of the 30~$\mu$m m waveguides. In subfigure (a): the circles correspond to a defect recording step of 3~$\mu$m, and the squares-4~$\mu$m.}
    \label{fig:Output_power_30um}
\end{figure}

\begin{figure}[htbp]
    \centering
    \begin{minipage}[b]{0.7\textwidth}
        \centering
        \includegraphics[width=\linewidth]{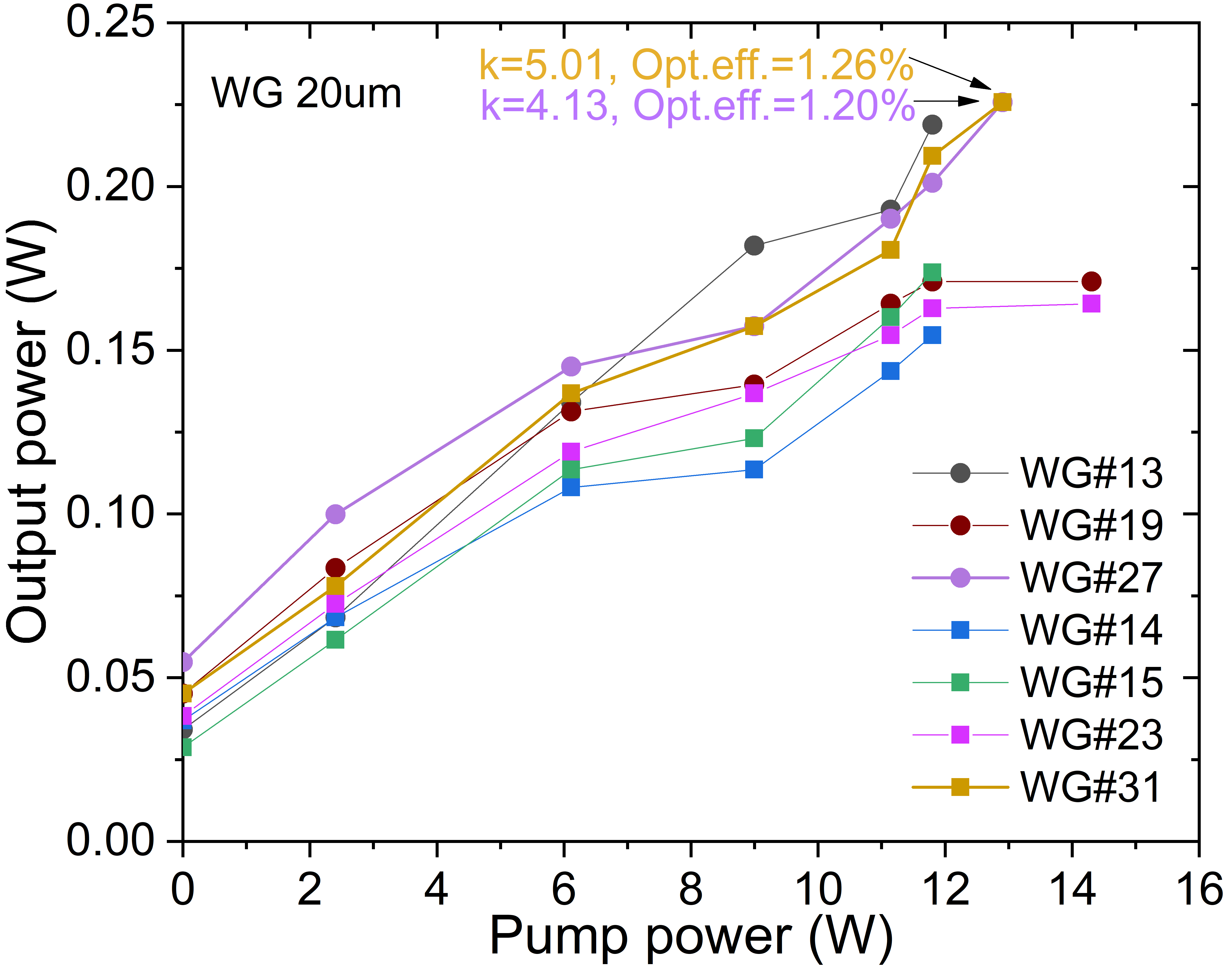} %
        \subcaption{Output power characteristics}
        \label{fig:20um_big_image}
    \end{minipage}
    \hfill
    \begin{minipage}[b]{0.27\textwidth}
        \centering
        \begin{subfigure}[b]{\textwidth}
            \centering
            \includegraphics[width=\linewidth]{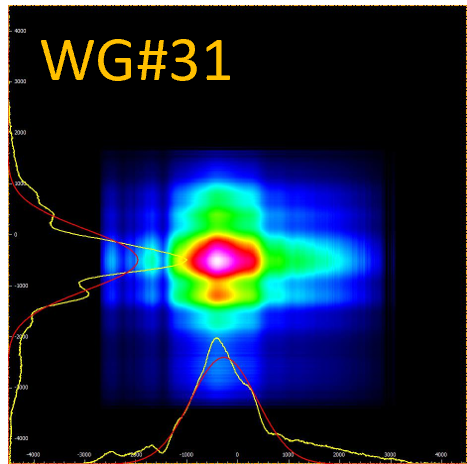} %
            \caption{WG\#31}
            \label{fig:20um_small_image1}
        \end{subfigure}
        \vfill
        \begin{subfigure}[b]{\textwidth}
            \centering
            \includegraphics[width=\linewidth]{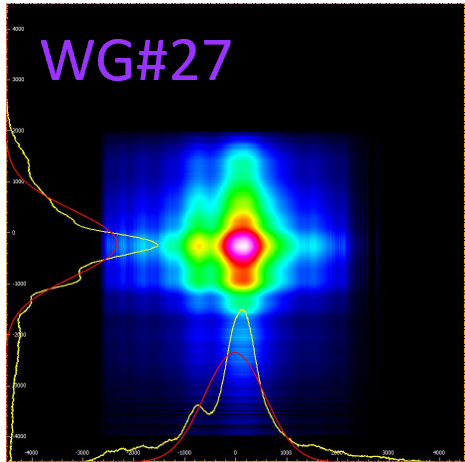} %
            \caption{WG\#27}
            \label{fig:20um_small_image2}
        \end{subfigure}
    \end{minipage}
    \caption{Output power characteristics (a) and beam profiles (b and c) of the 20~$\mu$m m waveguides. In subfigure (a): the circles correspond to a defect recording step of 3~$\mu$m, and the squares-4~$\mu$m.}
    \label{fig:Output_power_20um}
\end{figure}

Waveguides with diameters of 30 and 20~$\mu$m showed an order of magnitude lower results (see Figs.~\ref{fig:Output_power_30um}~and~\ref{fig:Output_power_20um} respectively). The average power levels were about 0.23-0.3~W and gain factors of 5-6.5 with optical efficiency of about 1.2-1.8\%. For both groups, high-order spatial modes are clearly visible, indicating mode leakage. Along with this, a low gain factor may also indicate mode leakage of the pump radiation beam. A more detailed discussion of the behavior of spatial modes in waveguides will be presented in the section with the results of theoretical simulation.

\section{Results of modelling}
\subsection{Model}\label{sec:model}
To explore the waveguiding properties of an amplifier, we based on the (2D+1)-dimensional model of laser amplification sketched in \cite{demesh2023threshold}. The amplifying and pumping complex fields $E(z,x,y)$ and $P(z,x,y)$ are distributed on a (x,y)-plane, which is transverse to a propagation axis $z$. One can write a combined system describing the system dynamics under the diffraction, spatially-graded refraction $n(x,y)$, and saturable gain in the following dimensionless form ($\Delta_{X,Y}$ is a transverse Laplacian describing diffraction in the paraxial approximation, the waveguiding potential is $\left( k_0/2 n_0 \right)\left( n_1^2 - n(x,y)^2 \right)\simeq k_0\left( n_1-n(x,y) \right)$):

\begin{align} 
\left[ \frac{1}{2} \Delta_{X,Y} - \left( X^2+Y^2\right)^m +i\left( \Lambda  - \sigma_r g_{max} \left( 1 - W_P \frac{\left| P(Z,X,Y) \right|^2}{1+V \left| E(Z,X,Y)) \right|^2}\right)\right) + i\frac{\partial }{\partial Z} \right]E(Z,X,Y)=0,\nonumber \\
    \left[ \frac{k_0}{2 k_0^\prime} \Delta_{X,Y} - \left[ \frac{k_0^\prime}{k_0}\left( X^2+Y^2\right)\right]^m + i\frac{\partial }{\partial Z} \right]P(Z,X,Y)= \label{eq:eq1}\\
    = -i \sigma_r g_{max}\left( 1 - W_P \frac{\left| P(Z,X,Y) \right|^2}{1+V \left| P(Z,X,Y)) \right|^2} \right) P(Z,X,Y). \nonumber
\end{align}

\noindent Transforming $E_{old}\exp{(-i V_0 z)} \to E_{new}$, where $V_0=\left(k_0/2 n_0\right)\left(n_1^2-n_0^2\right)$, defines new transverse and longitudinal scales for new variables $\left(X,Y,Z\right)$ \cite{haelterman1992dissipative,RAGHAVAN2000377,aschieri2011condensation}: $w_T=\sqrt[2 (m+1)]{w_0^{2 m}/k_0^2 n_0 \delta n}$ and $\zeta = k_0 n_0 w_T^2$. Here, $k_0$ and $n_0$ are the wavenumbers at the amplified wavelength and the corresponding cladding refractive index. $\delta n = n_1 - n_0$ is a graded refractive index contrast of waveguide ($n_1$ is the refractive index at the waveguide axis). We assume in Eq. (\ref{eq:eq1}) that the waveguiding properties for the pump $P$ are similar to those for $E$ with a correction taking into account the wavenumber $k^\prime$ for the pump. The waveguiding potential is assumed to be (super-)parabolic in shape with $m=1,2,3$. Since the seed pulse spectrum remains almost unchanged during the amplification process and the chirped pulse peak power remains small, we neglect the processes caused by nonlinear phase shift and group-delay dispersion. The $\Lambda$-parameter describes the integral spectral loss due to overlapping between the approximately rectangular pulse spectrum (width of $\approx $100 nm, Fig. 4b) centered at 2.35 $\mu$m and the Gaussian-like gain band (FWHM$\approx $800 nm \cite{6861943}). 

The saturable gain in Eq. (\ref{eq:eq1}) is described in the following way \cite{herrmann1987lasers}. The gain coefficient $g(x,y,z)$ for a four-level active medium can be defined in an quasi-two-level approximation through the pump, amplifying, and saturation ``velocities'' $\Phi_P$, $\Phi_A$ and $\Phi_S$, respectively, as:

\begin{align}
    g\left( x,y,z \right)=\frac{N_{Cr}\sigma_{em} \Phi_P(x,y,z)}{\Phi_P(x,y,z)+\Phi_A(x,y,z)+\frac{1}{T_r}}=\frac{g_0(x,y,z)}{1+\frac{\Phi_A(x,y,z)}{\Phi_S(x,y,z)}},\nonumber\\
    \Phi_P(x,y,z)=\sigma_a\frac{\left| P(x,y,z) \right|^2}{h \nu_p}, \Phi_A(x,y,z)=\sigma_{em}\frac{\left| E(x,y,z) \right|^2}{h \nu_{em}}, \label{eq:eq2}  \\ \Phi_S(x,y,z)=\Phi_P(x,y,z)+\frac{1}{T_r}.\nonumber
\end{align}

\noindent Here, $\sigma_{em}$ and $\sigma_a$ are the emission and absorption cross sections of Cr:ZnS active medium at 2.35 and 1.55 $\mu$m, respectively. $\nu_{em}$ and $\nu_a$ are the corresponding frequencies. $T_r$ is a gain relaxation time, and $N_{Cr}$ is a Cr-ions concentration. $g_0= g_{max} \Phi_P/\Phi_S$ is a small signal gain, and $g_{max}=N_{Cr} \sigma_{em}$ is a maximal gain. One has to note that the transition to the dimensionless gain coefficient in Eq. (\ref{eq:eq2}) implies a multiplication on $\zeta$. Squared amplitudes in Eq. (\ref{eq:eq2}) have the meaning of the corresponding intensities.

Under the reasonable assumption of $\Phi_S \approx 1/T_r$,
the gain in Eq. (\ref{eq:eq2}) can be defined as

\begin{equation}
    g\left( x,y,z \right)=\frac{g_0}{1+\left| E \right|^2/I_S}, \label{eq:eq3}
\end{equation}

\noindent where $I_S=h \nu_{em}/T_r \Sigma$ is a gain saturation intensity defining the intensity normalization in Eq. (\ref{eq:eq1}). $\Sigma$ is a signal beam area. 

The gain depletion in Eq. (\ref{eq:eq1}) is described by a formula for the ground-state occupation $N_1$ in a quasi-two-level system: $N_1=\left(g_{max} -g\right)/\left(\sigma_{em} + \sigma_a\right)$ so that $\sigma_r=\sigma_{em}/\left(\sigma_{em} + \sigma_a\right)$ in Eq. (\ref{eq:eq1}). 
As a result, we obtain Eq. (\ref{eq:eq1}) with $W_P = \sigma_a T_r/h \nu_a$ and $V = \sigma_{em} T_r/h \nu_{em}$.

\subsection{Results}

\begin{figure}
    \centering
    \includegraphics[scale=0.7]{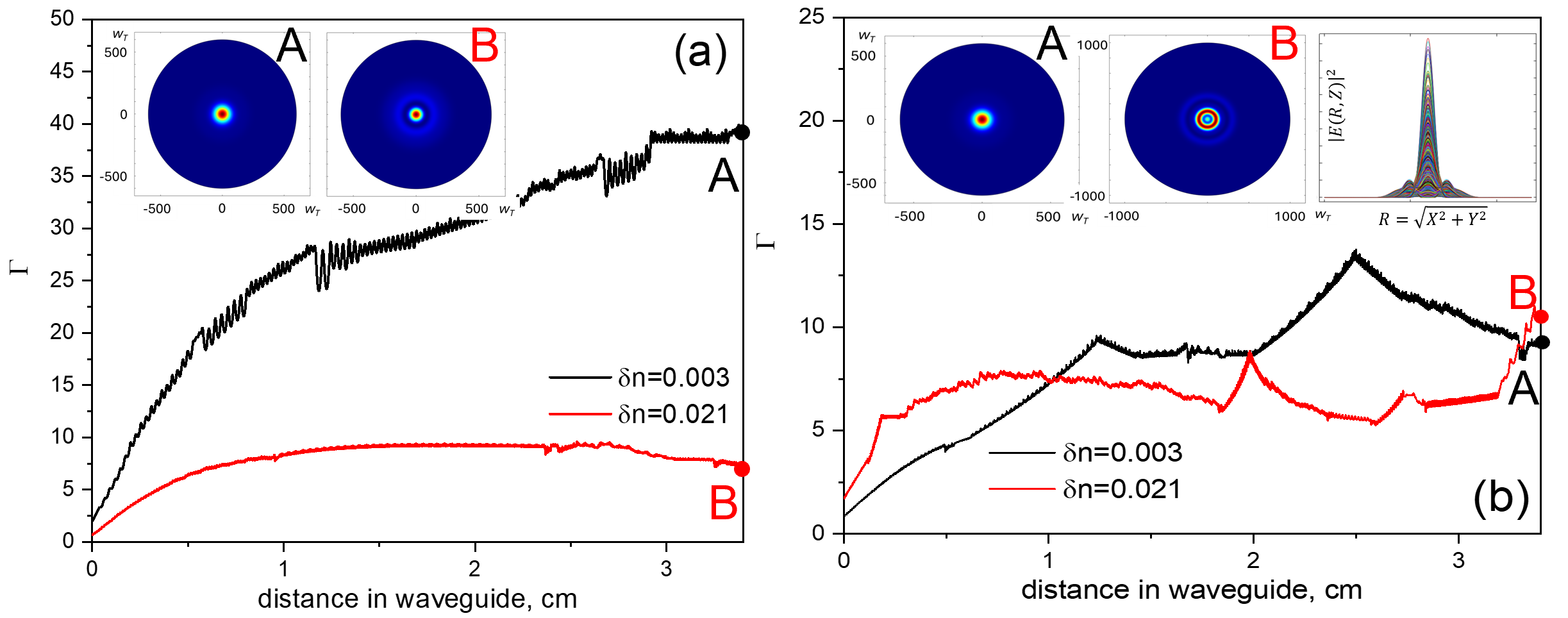}
    \caption{The amplification gain $\Gamma$ for $m$=1 $P(z=0,x,y)$ = 22 W (a) and 12 W (b). $w_0$= 50 (a) and 20 $\mu$m (b). $\delta n$ = 0.003 (black curves) and 0.021 (red curves). The left and central insets show the contour plots of the output mode $|E|^2$-profiles (points A and B). The right inset in (b) shows the mode profile averaged on $Z$ for $w_0$=20 $\mu$m, $\delta n$ = 0.003 (the colored filling corresponds to mode profiles at different $Z$ inside a waveguide).} 
    \label{fig:fig1th}
\end{figure}

\begin{figure}
    \centering
    \includegraphics[scale=0.7]{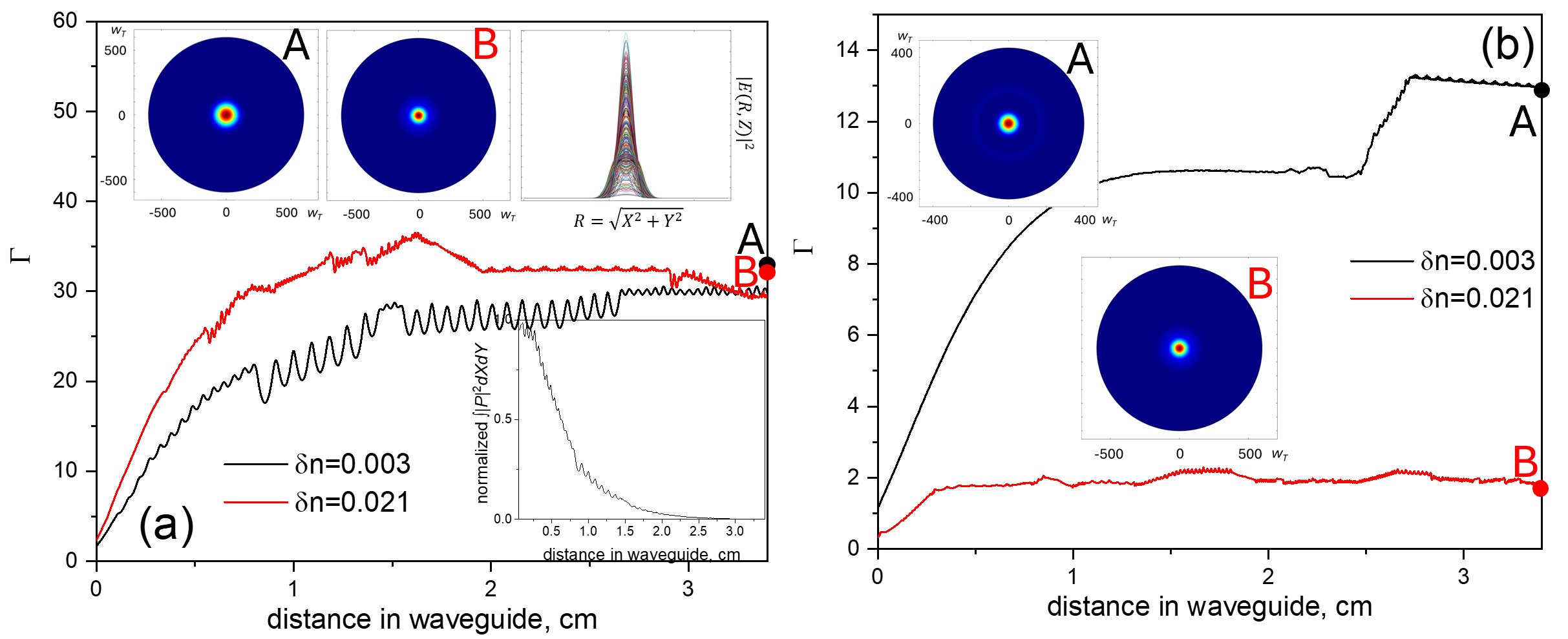}
    \caption{The same as in Fig. \ref{fig:fig1th} but for $m=2$. The right upper inset in (a) shows the mode profile averaged on $Z$ for $w_0$=50 $\mu$m, $\delta n$ = 0.021 (the colored filling corresponds to mode profiles at different $Z$ inside a waveguide). The bottom inset shows the normalized pump power $\int{|P(X,Y,Z)|^2 dX dY}$ vs. distance in a waveguide for $\delta n$=0.003.} 
    \label{fig:fig2th}
\end{figure}

\begin{figure}
    \centering
    \includegraphics[scale=0.7]{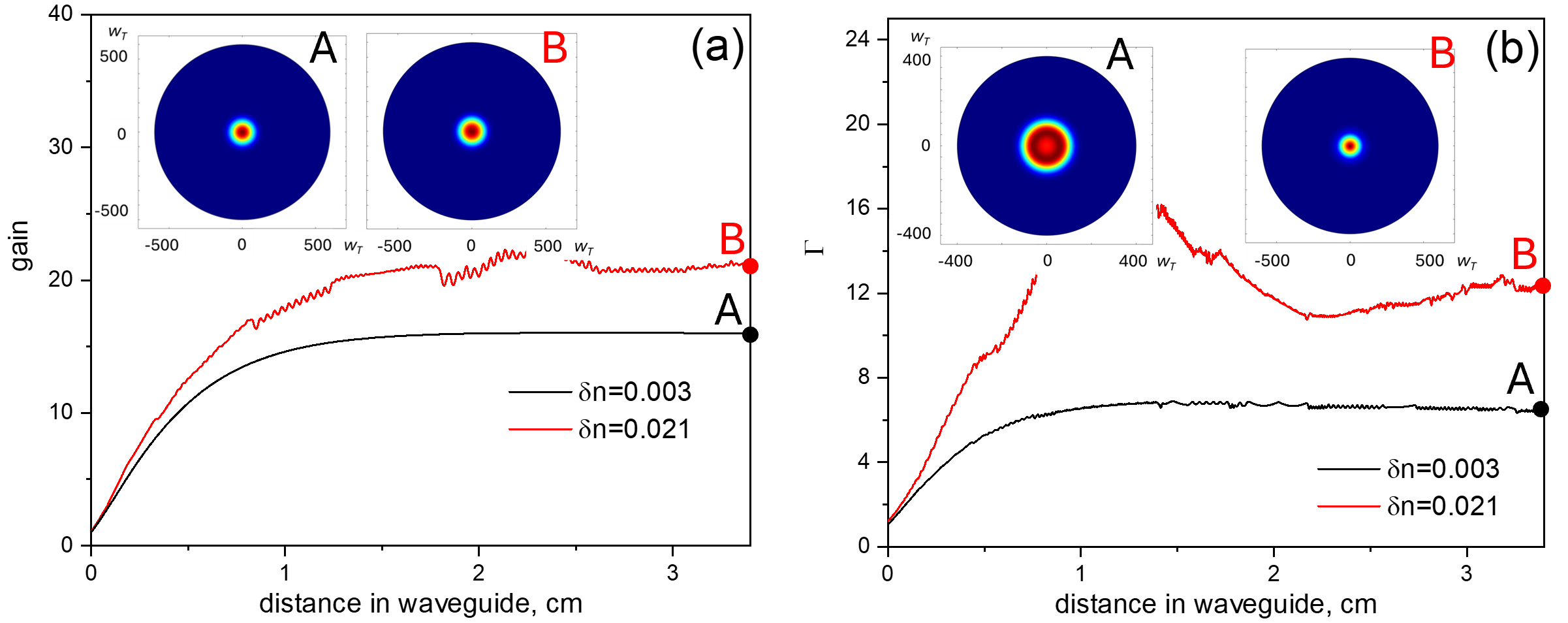}
    \caption{The same as in Fig. \ref{fig:fig1th} but for $m=3$.} 
    \label{fig:fig3th}
\end{figure}

We used the cylindrical geometry with Dirichlet's boundary conditions for the numerical simulations based on the model described in the previous Subsection \ref{sec:model} and used the COMSOL software. The results are shown in Figs. \ref{fig:fig1th}--\ref{fig:fig3th}. These Figures demonstrate the evolution of an effective gain $\Gamma$ over a distance $z$ along the waveguide axis. The effective gain of the amplified power is defined as

\begin{equation}\label{eq:G}
     \Gamma =\frac{\int_{-x_{min}}^{x_{max}}\int_{-y_{min}}^{y_{max}}\left|E\left(z,x,y;\, P(z,x,y)\right) \right|^2 dxdy}{\int_{-x_{min}}^{x_{max}}\int_{-y_{min}}^{y_{max}}\left|E\left(z_{out},x,y;\, P(z,x,y)=0\right) \right|^2 dxdy},
\end{equation}

\noindent where $E\left(z,x,y;\, P(z,x,y)\right)$ in the numerator defines an amplifying field evolving in parallel with the pump beam (see (\ref{eq:eq1})). The power $E\left(z_{out},x,y;\, P(z,x,y)=0\right)$ in the numerator corresponds to the field on the waveguide output ($z_{out}$=3.4 cm) after propagation without a pump. Both seed and pump are Gaussian at the input with equal beam sizes of 16 $\mu$m. We consider an input stretched seeding pulse of 2.94 W peak power and a rectangular spectrum of 100 nm width. Two waveguides are considered: with 50 and 20 $\mu$m sizes and pumped by 22 and 12 Watts, respectively. The waveguide slope indexes in (\ref{eq:eq1}) were $m=$1, 2, and 3.

As one can see from Fig. \ref{fig:fig1th} (a, black curve), the multimode nature of the 50 $\mu$m waveguide causes the inter-mode beatings for $\Delta n$=0.003\footnote{The waveguide size $w_0$ providing a single mode propagation is of $\approx$8 $\mu$m in this case for $\lambda$=2.35 $\mu$m from $\mathcal{V}=(2\pi w_0/\lambda) \sqrt{n_1^2-n_9^2}<2.4$ \cite{snyder1983optical}.}. As a result, the $\Gamma$-coefficient oscillates.

Such oscillations are reduced for $\Delta n$=0.021 (Fig. \ref{fig:fig1th} (a), red curve) by the cost of the irregularity growth due to squeezing a zero-mode size down to $\approx$2.4 $\mu$m (the mode sizes of both seed and pump Gaussian input beams equal 16 $\mu$m)\footnote{A zero-mode size $\approx 2 w_0/\sqrt{2 (\mathcal{V}^2/2-1)}$.}. The resulting excitation of high-order modes is visible in Fig. \ref{fig:fig1th} (a, inset B). The irregularity of overlapping between the amplifying and pump modes decreases $\Gamma$. 


Decreasing $w_0$ (Fig. \ref{fig:fig1th} (b)) reduces $\Gamma$. Since the waveguide squeezing increases the mode overlapping, it leads to stronger inter-modal crosstalk and enhances the inter-modal coupling. As a result, the gain decreases. The reduced input pump power and the mode volume also contribute to $\Gamma$-lowering.
$\Delta n$-growth intensifies the inter-modal crosstalk and lowers the mode volume due to stronger confinement. Both cause a further gain reduction (Fig. \ref{fig:fig1th} (b, red curve)). The excitation of higher-order modes is clearly visible in the output mode profiles (Fig. \ref{fig:fig1th} (b, left and central insets)) and the averaged mode profile (Fig. \ref{fig:fig1th} (b, right inset)). 


With increasing index contrast ($m$=2 in Fig. \ref{fig:fig2th}), the waveguide supports fewer modes. The inter-mode coupling decreases, and the quality of the output mode improves (upper insets in Fig. \ref{fig:fig2th} (a)). One can see from Fig. \ref{fig:fig2th} (a) that the gain $\Gamma$ is close for both $\Delta n$=0.003 and 0.021, but the oscillation frequency is higher and its amplitude lower for the latter. The amplification remains reduced for $w_0$=20 $\mu$m, especially for high $\Delta n$ (Fig. \ref{fig:fig2th} (b)). 

The situation becomes more interesting with a further $m$-increase, that is, the transition to a sharper $n$-gradient (Fig. \ref{fig:fig3th}). For $w_0$=50 $\mu$m, the amplification with the high-amplitude and low-frequency oscillations of $\Gamma$ for $\Delta n$=0.021 dominates over that for $\Delta n$=0.003 (Fig. \ref{fig:fig3th} (a)). Also, one can see such domination for $w_0$=20 $\mu$m (Figure (b)). However, the higher modes worsen the output mode quality for the latter (inset A). Namely, the growth of the higher-order modes with $\Delta n$ (roughly $\propto \mathcal{V}^2$) could explain the $\Gamma$-rise due to their ``harvesting'' (red curves versus black ones in Fig. \ref{fig:fig3th}), when a confining potential approaches step-index one.

\section{Discussion}

The experimental results show that the defect recording step does not have a significant influence on the waveguide performance in our particular case. 

Analyzing the output beam profiles for the 50 and 40~$\mu$m waveguides, as well as the results of M$^{2}$ factor measurements, it is clear that the spatial profile formation of the amplified pulses beam occurs under the spatial mode spectrum unique for the particular waveguide. However, it can be assumed that, unlike fiber lasers, waveguides have a clear and rigidly defined geometry (and short length), which makes the main factor influencing the quality of the output beam formation just the coupling conditions of the incident radiation together with the spectrum of the waveguide’s eigenmodes. Finally, this makes obtaining high spatial quality of the output beam profile possible even for large-diameter waveguides, which are multimode in nature.

We should note that the waveguide excludes the free-space sectors in an amplifier and eliminates the gain band narrowing due to atmospheric water at $\gtrsim$2.4 $\mu$m. Thus, the broad gain band can effectively amplify spectrally broad pulses. Thus, our main claim is using a solid-state waveguide excluding the free-space sectors in an amplifier that allows, in particular, eliminating the gain band narrowing due to atmospheric water at $\gtrsim$2.4 $\mu$m. Thus, the broad gain band can effectively amplify spectrally broad pulses. 

Our experiment demonstrates explicitly a realization of waveguiding, which is especially visible in Fig. \ref{fig:visible_waveguide}. Although the field intensities were insufficient for realizing the spatial mode self-cleaning \cite{krupa2017spatial}, one may see from this Figure a guided fourth-order harmonic demonstrating an effective higher-order spatial mode confinement. The theoretical analysis shows that the multimode nature of the waveguides considered impacts both the gain and the transverse mode profile depending on the waveguide confining potential profile and depth (Figs. \ref{fig:fig1th}--\ref{fig:fig3th}). 

For a slight refractive index contrast $\delta$=0.003 and a large waveguide size $w_0$=50 $\mu$m, the gain coefficient $\Gamma$ decreases by approaching the refractive index profile to a step-shaped one with the $m$-growth (left pictures in Figs. \ref{fig:fig1th}--\ref{fig:fig3th}, black curves). That reduces the number of excited modes and their volume. Simultaneously, the overlapping of modes decreases, suppressing the modes' crosstalk and oscillations (Fig. \ref{fig:fig3th}, a, black curve).    

Reduction of the waveguide size $w_0$ decreases $\Gamma$ (Figs. \ref{fig:fig1th}--\ref{fig:fig3th}, b) due to lowering of the mode numbers and pump power. The last is necessary to avoid the waveguide damage. Simultaneously, the difference in the gain coefficients for $w_0$= 50 and 20 $\mu$m decreases with the $m$-growth. Also, one may see a total $\Gamma$-decrease with $m$-growth, resulting from a mode spectrum ``thinning'', simultaneously diminishing the modes' crosstalk.

Qualitatively, the simulations reproduce the experimental results. Some quantitative disagreement in a $\Gamma$ value for $w_0$=50 $\mu$m could be explained in the following way. A circular symmetry assumed in the model results in the larger mode volumes and the overlapping between them than in the near-rectangular waveguides used in the experiment (e.g., see Fig. \ref{fig:Output_power_50um}). That could enhance the modes' crosstalk and earlier gain depletion (bottom inset in Fig. \ref{fig:fig2th}). One has to note that a perfect pump depletion in a waveguide agrees with the experimental observations.

\begin{figure}
    \centering
    \includegraphics[scale=0.7]{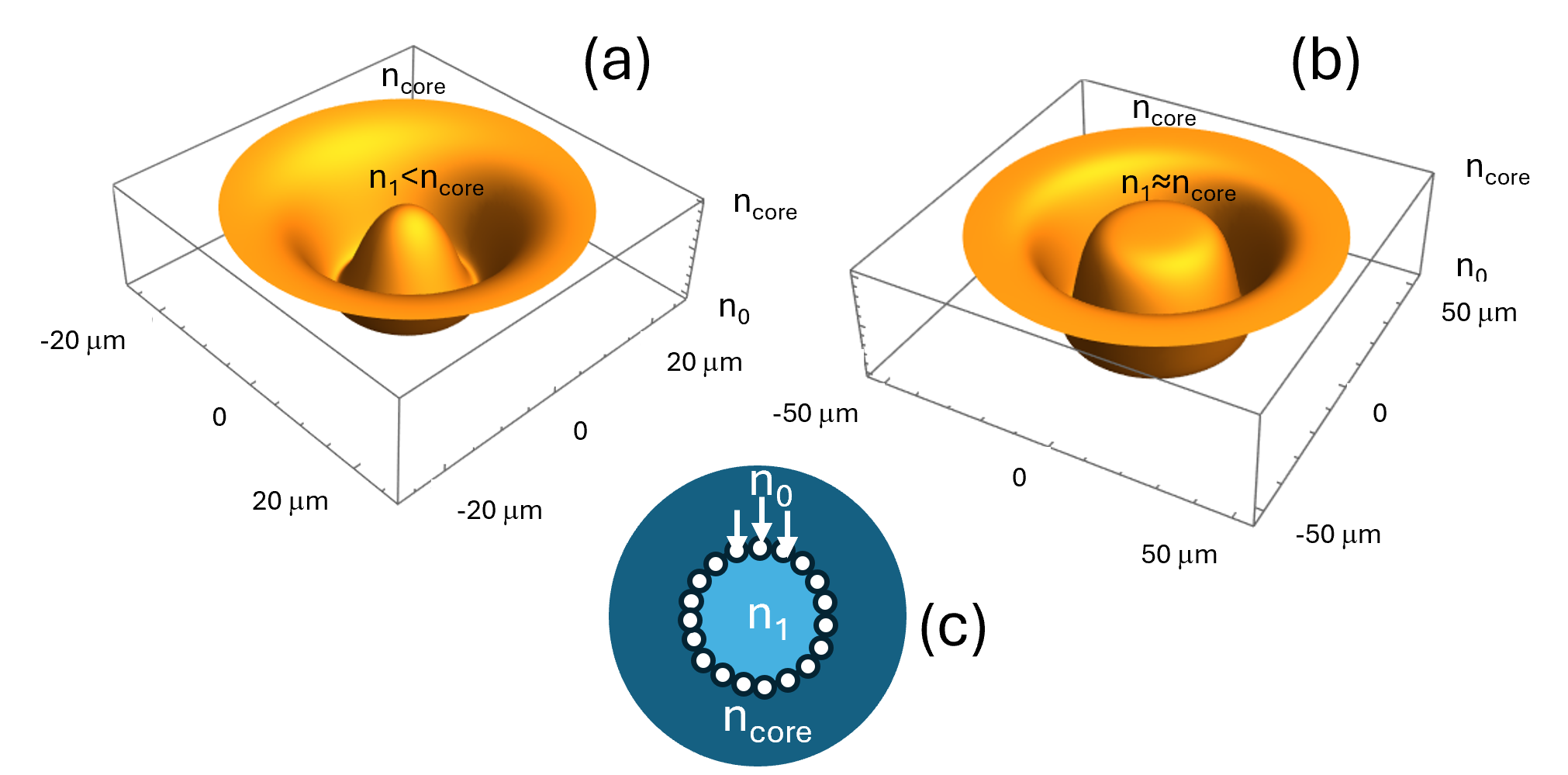}
    \caption{Schematic presentation of the refractive index profiles for the thin (a) and thick (b) waveguides. The holes (shown by the white circles in (c)) are manufactured by laser processing inside a bulk ZnS-sample with the refractive index $n_{core}$. The modified refractive index inside the holes is $n_0$ and rises to $n_1$ at the waveguide axis. Since the length of refractive index relaxation is about 10 $\mu$m, the effective $\delta n$ would be lower for a thin waveguide, and $m$ is larger for a thick one.}
    \label{fig:fig_waveguide}
\end{figure}

An additional factor is that the waveguides used in the experiment are, in fact, the large mode area photonic crystal waveguides \cite{limpert2004low,dong2015advanced}, which are shown schematically in Fig. \ref{fig:fig_waveguide}. As a result, their mode structure depends on the large-scale (a, b) and small-scale (c) geometries. That provides a broad range of tools for mode manipulation, including power scaling by mode area and mode synthesis. In our case, one could address a transition from a thin waveguide to a thick one in a parallel with $\delta n$ and $m$ growth. Simultaneously, the upper right insets in Figs. \ref{fig:fig1th},\ref{fig:fig2th} demonstrates a decreasing contribution of higher-order modes to the averaged mode profile (i.e., $M^2$-factor approaching 1) for such a transition that corresponds to the experimental results.    

Given the adequate description of the waveguides, one can optimize the design and writing procedure. The ultrashort-pulse modification approach, especially the single-pulse modification possibility, allows nearly free-form designs limited only by programming skills. By controlling the pulse energy during the writing procedure, one can change the waveguide form and shape along the propagation axis (tapering) as well as the index contrast.  

\section{Conclusion and outlook}
We demonstrate that using a large cross-section multimode Cr$^{2+}$:ZnS waveguide allows reaching the gain factors up to~75 in a single pass. To our knowledge, this is the first demonstration of the waveguide amplifier in the crystalline active medium operating in a single transverse mode regime. Experiments and the results of our theoretical analysis show the interplay between the non-dissipative (transversely graded refractive index) and dissipative (graded gain) confinement potentials. This results in integrating higher-order modes into lower-order ones (spatial mode synthesis or ``mode condensation'' \cite{krupa2017spatial}), potentially allowing effective use of the whole gain volume and energy harvesting in the compact multimode waveguides. As an outlook, one may anticipate realizing a high-repetition-rate waveguide self-mode-locked oscillator \cite{demesh2023threshold} and a photonic crystal-like amplifier in a laser-processed Cr$^{2+}$:ZnS bulk medium. All this opens up a totally new avenue towards an integrated format and power scaling of ultra-broadband and ultra-short pulsed waveguide lasers – a new laser technology that can be compared only to fiber lasers with the advantages of crystalline solid-state lasers.

\begin{backmatter}
\bmsection{Funding} Norges Forskningsråd (\#303347 (UNLOCK), \#326241 (Lammo-3D), \#326503 (MIR)); ATLA Lasers AS.

\bmsection{Acknowledgments}
The work of AR, VLK, MD and ITS was supported by NFR projects \#303347 (UNLOCK),
\#326503 (MIR), \#326241 (Lammo-3D) and by ATLA Lasers AS.

\bmsection{Disclosures}  ES: ATLA Lasers AS (I), ITS: ATLA Lasers AS (I, S).

\bmsection{Data Availability Statement}
Data underlying the results presented in this paper are not publicly available at this time but may be obtained from the authors upon reasonable request.

\end{backmatter}

\bibliography{Bibliography}

\end{document}